\documentclass[11pt]{article}
\usepackage{fullpage,epsf,amssymb,amsthm,amsfonts,amsmath,latexsym, graphicx,array,extarrows,mathtools,mathstyle,mathrsfs}
\usepackage[normalem]{ulem}

\newcommand{\overbar}[1]{\mkern 1.5mu\overline{\mkern-1.5mu#1\mkern-1.5mu}\mkern 1.5mu}

\let\oldsqrt\sqrt
\def\sqrt{\mathpalette\DHLhksqrt}
\def\DHLhksqrt#1#2{%
\setbox0=\hbox{$#1\oldsqrt{#2\,}$}\dimen0=\ht0
\advance\dimen0-0.2\ht0
\setbox2=\hbox{\vrule height\ht0 depth -\dimen0}
{\box0\lower0.4pt\box2}}

\numberwithin{equation}{section}

\title{\bf{The non-perturbative structure of the photon and gluon propagators}} 
\author{\large{Peter Lowdon} \\
\ \\
\textit{\small{SLAC National Accelerator Laboratory, 2575 Sand Hill Rd, Menlo Park, CA 94025, USA}} \\
\textit{\small{E-mail: lowdon@slac.stanford.edu}}}
\date{}
\begin{document}
\begin{flushright} SLAC-PUB-16924   \end{flushright}
\vspace{15mm} 
{\let\newpage\relax\maketitle}
\setcounter{page}{1}
\pagestyle{plain}

\abstract 

\noindent 
The non-perturbative structure of the photon and gluon propagators plays an important role in governing the dynamics of quantum electrodynamics (QED) and quantum chromodynamics (QCD) respectively. Although it is often assumed that these interacting field propagators can be decomposed into longitudinal and transverse components, as for the free case, it turns out that in general this is not possible. Moreover, the non-abelian gauge symmetry of QCD permits the momentum space gluon propagator to contain additional singular terms involving derivatives of $\delta(p)$, the appearance of which is related to confinement. Despite the possibility of the failure of the transverse-longitudinal decomposition for the photon and gluon propagators, and the appearance of singular terms in the gluon propagator, the Slavnov-Taylor identity nevertheless remains preserved.

\newpage

\section{Introduction \label{section1}}

Correlators, and thus propagators, are the central objects of interest in any quantum field theory (QFT). Despite their importance, the non-perturbative structure of propagators in physical theories such as quantum electrodynamics (QED) and quantum chromodynamics (QCD) remains largely unknown. Nevertheless, there are several techniques which have the potential to probe this non-perturbative behaviour. Axiomatic quantum field theory (AQFT) is one such approach, and consists of defining a QFT in a mathematically rigorous manner via the definition of a series of physically motivated axioms~\cite{Streater_Wightman64,Haag96,Nakanishi_Ojima90,Strocchi13,Bogolubov_Logunov_Oksak90}. Although different axiomatic schemes have been proposed, these schemes generally consist of a common core set of axioms which are often referred to as the \textit{Wightman axioms}~\cite{Streater_Wightman64}. These axioms include assumptions such as relativistic covariance, fields as (operator-valued) distributions, and locality\footnote{See~\cite{Streater_Wightman64,Haag96,Nakanishi_Ojima90,Strocchi13,Bogolubov_Logunov_Oksak90} for a more in-depth discussion of these axioms.}. \\

\noindent
In the case of quantised gauge theories such as QED and QCD, the standard Wightman axioms no longer apply. In particular, gauge symmetry provides an obstacle to the locality of fields in the theory. To quantise a gauge theory one therefore has to either accept that fields can be non-local, as is the case in Coulomb gauge, or one can preserve locality by adopting a \textit{local quantisation}. In local quantisations, additional degrees of freedom are introduced into the theory, resulting in a space of states $\mathcal{V}$ which no longer possesses a positive-definite inner product. Since negative norm states are unphysical, one must define an external condition in order to specify the physical states $\mathcal{V}_{\text{phys}} \subset \mathcal{V}$. For gauge theories such as QED and QCD, \textit{BRST quantisation} is an important example of a local quantisation. In this case, auxiliary gauge-fixing and ghost term are added to the equations of motion of the theory in order to break the gauge invariance, and thus preserve the locality of the fields. Although the gauge-fixed theory is no longer gauge invariant, it remains invariant under a residual Becchi-Rouet-Stora-Tyutin (BRST) symmetry, which has a corresponding conserved charge $Q_{B}$. Physical states are then defined by the requirement that the quantised equations of motion must hold for these states, and it turns out that this is equivalent to the condition: $Q_{B}\mathcal{V}_{\text{phys}}=0$~\cite{Strocchi13}. Due to the preservation of locality, BRST quantisation is usually employed when analysing the non-perturbative structure of the photon and gluon propagators. The modification of the Wightman axioms required to facilitate the indefinite inner product space of states $\mathcal{V}$ in this approach is referred to as the Pseudo-Wightman formalism~\cite{Bogolubov_Logunov_Oksak90}. Although many of the results derived from the standard Wightman axioms are maintained in this formalism~\cite{Strocchi78}, the modification of the axioms can lead to significant changes in the structure of correlators and propagators, and it is precisely these differences which will be explored in this paper.  \\

\noindent
The rest of this paper is structured as follows: in Sec.~\ref{np_gen} the general properties of Lorentz covariant correlators is outlined, and these properties are applied in order to derive the general form of the correlator and propagator of an arbitrary vector field; in Sec.~\ref{vector_examples} the results derived in Sec.~\ref{np_gen}, together with the model-dependent constraints, are used to derive the structure of the non-perturbative photon propagator in free (quantised) electromagnetism and QED, as well as the gluon propagator in QCD; in Sec.~\ref{tran_long}, the issue of whether a transverse-longitudinal decomposition exists for the interacting photon and gluon propagator is discussed; and finally in Sec.~\ref{concl} the key findings are summarised.

\section{The non-perturbative structure of vector correlators and propagators}
\label{np_gen}

\subsection{The vector correlator}
\label{vect_corr_sect}

In axiomatic formulations of QFT~\cite{Streater_Wightman64}, the basic field correlators $\langle 0 | \phi_{1}(x_{1})\phi_{2}(x_{2}) |0\rangle = T_{(1,2)}(x_{1}-x_{2})$ are defined to be tempered distributions $\mathcal{S}'(\mathbb{R}^{1,3})$, and hence their Fourier transforms $\widehat{T}_{(1,2)}(p) = \mathcal{F}\left[T_{(1,2)}(x_{1}-x_{2})\right]$ are in $\mathcal{S}'(\mathbb{R}^{1,3})$. Moreover, since quantised fields are also assumed to transform covariantly under Lorentz transformations, $\widehat{T}_{(1,2)}(p)$ is a \textit{Lorentz covariant} distribution, and therefore satisfies the following condition~\cite{Bogolubov_Logunov_Oksak90}: 
\begin{align}
\widehat{T}_{(1,2)}(\Lambda p) = S(\Lambda) \, \widehat{T}_{(1,2)}(p),
\end{align}     
where $\Lambda \in \overbar{\mathscr{L}_{+}^{\uparrow}} \cong \mathrm{SL}(2,\mathbb{C})$. The structure of the Lorentz covariant distribution $\widehat{T}_{(1,2)}(p)$ is dependent upon how the fields $\phi_{1}$ and $\phi_{2}$ transform under Lorentz transformations. In particular, $\widehat{T}_{(1,2)}(p)$ has the following decomposition~\cite{Bogolubov_Logunov_Oksak90}:
\begin{align}
\widehat{T}_{(1,2)}(p) = \sum_{\alpha=1}^{\mathscr{N}}Q_{\alpha}(p) \, \widehat{T}_{\alpha (1,2)}(p),
\label{decomp_cov}
\end{align}    
where $\widehat{T}_{\alpha (1,2)}(p)$ are \textit{Lorentz invariant} distributions [i.e. $\widehat{T}_{\alpha (1,2)}(\Lambda p) = \widehat{T}_{\alpha (1,2)}(p)$], and $Q_{\alpha}(p)$ are Lorentz covariant polynomial functions of $p$ which carry the Lorentz index structure of $\phi_{1}$ and $\phi_{2}$. Before discussing the specific structure of the photon and gluon correlators and propagators, one must first consider the general case where $\phi_{i}$ are both arbitrary vector fields. Given that $\phi_{1}=A_{\mu}$ and $\phi_{2}=A_{\nu}$, it turns out that there are two possible Lorentz covariant polynomials: $Q_{1}(p) = g_{\mu\nu}$ and $Q_{2}(p) = p_{\mu}p_{\nu}$. Due to Eq.~(\ref{decomp_cov}) it therefore follows that
\begin{align}
\widehat{D}_{\mu\nu}(p) &= \mathcal{F}\left[\langle 0| A_{\mu}(x)A_{\nu}(y)|0\rangle \right] =  g_{\mu\nu} \, \widehat{D}_{1}(p) + p_{\mu}p_{\nu} \, \widehat{D}_{2}(p).
\label{vector_decomp}
\end{align}  
\ \\
\noindent
In order to further specify the structure of $\widehat{D}_{\mu\nu}(p)$ one must first understand the behaviour of the Lorentz invariant components $\widehat{D}_{1}(p)$ and $\widehat{D}_{2}(p)$. It is well known that Lorentz invariant distributions $\widehat{T}_{\alpha} \in \mathcal{S}'(\mathbb{R}^{1,3})$ have certain structural properties. In particular, if $\widehat{T}_{\alpha}$ is restricted to have support in the closed forward light cone $\overbar{V}^{+}$, as is required in axiomatic formulations of QFT, $\widehat{T}_{\alpha}$ can be written in the following general manner~\cite{Bogolubov_Logunov_Oksak90}:   
\begin{align}
\widehat{T}_{\alpha}(p) =  P(\partial^{2})\delta(p) + \int_{0}^{\infty} ds \, \theta(p^{0})\delta(p^{2}-s) \rho_{\alpha}(s), 
\label{KL_gen_rep} 
\end{align}   
where $P(\partial^{2})$ is some arbitrary polynomial of finite order in the d'Alembert operator $\partial^{2} = g_{\mu\nu}\frac{\partial}{\partial p_{\mu}}\frac{\partial}{\partial p_{\nu}}$ (with complex coefficients), and $\rho_{\alpha}(s) \in \mathcal{S}'(\overbar{\mathbb{R}}_{+})$. This is the \textit{spectral representation} of $\widehat{T}_{\alpha}$, and $\rho_{\alpha}$ is the \textit{spectral density}. In the case of the vector field correlator (Eq.~(\ref{vector_decomp})), Eq.~(\ref{KL_gen_rep}) can be used to write $\widehat{D}_{\mu\nu}(p)$ in the form
\begin{align}
\widehat{D}_{\mu\nu}(p) &=    \int_{0}^{\infty} ds \, \theta(p^{0})\delta(p^{2}-s) \left[ g_{\mu\nu}\rho_{1}(s) + p_{\mu}p_{\nu} \rho_{2}(s) \right]  + \left[g_{\mu\nu}P_{1}(\partial^{2}) + p_{\mu}p_{\nu}P_{2}(\partial^{2}) \right]\delta(p),
\label{vector_decomp_further}
\end{align}
where $P_{1}$ and $P_{2}$ are polynomials of finite order. Performing the inverse Fourier transform of this expression leads to the following general representation of the position space correlator:
\begin{align*}
\langle 0| A_{\mu}(x)A_{\nu}(y)|0\rangle &=  \frac{i}{2\pi}\int_{0}^{\infty} ds \big[ -g_{\mu\nu}\rho_{1}(s) +\rho_{2}(s) \partial_{\mu}\partial_{\nu}  \big]D^{(-)}(x-y;s) \nonumber \\
& \hspace{5mm}  +\frac{1}{(2\pi)^{4}}  \left[g_{\mu\nu}P_{1}\left(-(x-y)^{2}\right)  -\partial_{\mu}\partial_{\nu} P_{2}\left(-(x-y)^{2}\right)\right].
\end{align*}
$P_{1}$ and $P_{2}$ are arbitrary complex polynomials of finite order and hence one can set: $P_{1}(\partial^{2}) = \sum_{l=0}^{L}a_{l}(\partial^{2})^{l}$, and $P_{2}(\partial^{2}) = \sum_{m=0}^{M}b_{m}(\partial^{2})^{m}$ where $a_{l},b_{m} \in \mathbb{C}$. Since the polynomial term $P_{2}\left(-(x-y)^{2}\right)$ involves derivatives, not all of the terms will contribute to the correlator. In fact, one can write
\begin{align*}
\partial_{\mu}\partial_{\nu} P_{2}\left(-(x-y)^{2}\right) =  -2b_{1}g_{\mu\nu} + \partial_{\mu}\partial_{\nu} \underbrace{\left(\sum_{m=2}^{M}b_{m}(-(x-y)^{2})^{m}  \right)}_{:= \widetilde{P}_{2}\left(-(x-y)^{2}\right) }. 
\end{align*}    
Finally, by setting $\tilde{a}_{0} = a_{0} + 2b_{1}$ ($\tilde{a}_{l} = a_{l}$ for $l \geq 1$)   
\begin{align}
\langle 0| A_{\mu}(x)A_{\nu}(y)|0\rangle  &= \frac{i}{2\pi}\int_{0}^{\infty} ds \left[ -g_{\mu\nu}\rho_{1}(s) +\rho_{2}(s) \partial_{\mu}\partial_{\nu}  \right]D^{(-)}(x-y;s) \nonumber \\
&\hspace{10mm} + \frac{1}{(2\pi)^{4}}\left[g_{\mu\nu}\widetilde{P}_{1}\left(-(x-y)^{2}\right) -\partial_{\mu}\partial_{\nu} \widetilde{P}_{2}\left(-(x-y)^{2}\right) \right],
\label{vector_decomp_pos}
\end{align}     
where now $\widetilde{P}_{1}(-(x-y)^{2}) = \sum_{l=0}^{L}\tilde{a}_{l}(-(x-y)^{2})^{l}$. 

\subsection{The vector propagator}

In general, the vector propagator involves a time-ordered product of fields, and is defined as:
\begin{align}
\langle 0|T\{ A_{\mu}(x)A_{\nu}(y)\}|0\rangle : &= \theta(x^{0}-y^{0}) \langle 0| A_{\mu}(x)A_{\nu}(y)|0\rangle + \theta(y^{0}-x^{0})\langle 0| A_{\nu}(y)A_{\mu}(x)|0\rangle.
\label{t_ordered_expl}
\end{align}
Using the spectral representation of the vector correlator in Eq.~(\ref{vector_decomp_further}), the propagator can be written
\begin{align}
\langle 0|T\{ A_{\mu}(x)A_{\nu}(y)\}|0\rangle &=  \theta(x^{0}-y^{0}) \int_{0}^{\infty} ds \int \frac{d^{4}p}{(2\pi)^{4}} \, e^{-ip(x-y)}\theta(p^{0})\delta(p^{2}-s) \left[ g_{\mu\nu}\rho_{1}(s) + p_{\mu}p_{\nu} \rho_{2}(s) \right] \nonumber \\
& \hspace{3mm} + \theta(x^{0}-y^{0})\int \frac{d^{4}p}{(2\pi)^{4}} \, e^{-ip(x-y)} \left[g_{\mu\nu}\widetilde{P}_{1}(\partial^{2}) + p_{\mu}p_{\nu}\widetilde{P}_{2}(\partial^{2}) \right]\delta(p) \nonumber \\
& \hspace{3mm} + \theta(y^{0}-x^{0}) \int_{0}^{\infty} ds \int \frac{d^{4}p}{(2\pi)^{4}} \, e^{ip(x-y)}\theta(p^{0})\delta(p^{2}-s) \left[ g_{\mu\nu}\rho_{1}(s) + p_{\mu}p_{\nu} \rho_{2}(s) \right] \nonumber \\
& \hspace{3mm} + \theta(y^{0}-x^{0})\int \frac{d^{4}p}{(2\pi)^{4}} \, e^{ip(x-y)} \left[g_{\mu\nu}\widetilde{P}_{1}(\partial^{2}) + p_{\mu}p_{\nu}\widetilde{P}_{2}(\partial^{2}) \right]\delta(p).
\label{t_ordered} 
\end{align}
In order to simplify this expression one can use the relation
\begin{align*}
\partial_{\mu}^{x}\partial_{\nu}^{x}\left[\theta(x^{0}-y^{0})e^{-ip(x-y)} + \theta(y^{0}-x^{0})e^{ip(x-y)}   \right] = & -p_{\mu}p_{\nu}\left[\theta(x^{0}-y^{0})e^{-ip(x-y)} + \theta(y^{0}-x^{0})e^{ip(x-y)}   \right] \\
& -i (p_{\mu}g_{\nu 0} + p_{\nu}g_{\mu 0}) \, \delta(x^{0}-y^{0})\left[ e^{-ip(x-y)} + e^{ip(x-y)} \right] \\
& +g_{\mu 0}g_{\nu 0} \, \delta'(x^{0}-y^{0}) \left[ e^{-ip(x-y)} - e^{ip(x-y)} \right],
\end{align*} 
which upon substitution into Eq.~(\ref{t_ordered}) implies that the vector propagator has the following general structure:
\begin{align}
\langle 0|T\{ A_{\mu}(x)A_{\nu}(y)\}|0\rangle &= \int_{0}^{\infty} \frac{ds}{2\pi} \, \left[ -g_{\mu\nu}\rho_{1}(s) + \rho_{2}(s)\partial_{\mu}^{x}\partial_{\nu}^{x} \right]i\Delta_{F}(x-y;s) \nonumber \\
& \hspace{15mm}-\frac{i}{2\pi}g_{\mu 0}g_{\nu 0} \, \delta(x-y) \int_{0}^{\infty} ds \, \rho_{2}(s) \nonumber \\
& \hspace{15mm} +\frac{1}{(2\pi)^{4}} \left[ g_{\mu\nu}\widetilde{P}_{1}\left(-(x-y)^{2}\right)-\partial_{\mu}^{x}\partial_{\nu}^{x} \widetilde{P}_{2}\left(-(x-y)^{2}\right) \right],
\label{general_propagator_pos}
\end{align} 
and thus the Fourier transformed propagator $\widehat{D}_{\mu\nu}^{F} = \mathcal{F}\left[\langle 0|T\{ A_{\mu}(x)A_{\nu}(y)\}|0\rangle \right]$ is given by
\begin{align}
\widehat{D}_{\mu\nu}^{F}(p) &=   i\int_{0}^{\infty} \frac{ds}{2\pi} \, \frac{\left[ g_{\mu\nu}\rho_{1}(s) + p_{\mu}p_{\nu}\rho_{2}(s) \right]}{p^{2}-s +i\epsilon} -\frac{i}{2\pi}g_{\mu 0}g_{\nu 0} \int_{0}^{\infty} ds \, \rho_{2}(s)  \nonumber \\
&\hspace{15mm} +g_{\mu\nu}\widetilde{P}_{1}(\partial^{2})\delta(p)+ p_{\mu}p_{\nu} \widetilde{P}_{2}(\partial^{2})\delta(p).
\label{general_propagator_mom}
\end{align}
A shared feature of the position and momentum space vector propagators is that they both contain an explicitly non-covariant term proportional to $g_{\mu 0}g_{\nu 0}$. This is in fact not surprising because unlike correlators, propagators involve time-ordered fields, and this requires one to single out a non-covariant plane ($x^{0}-y^{0}=0$) with which to chronologically order the fields. It is clear from Eq.~(\ref{general_propagator_mom}) that whether or not this non-covariant term appears depends on the integral of the spectral density $\rho_{2}$. \\

\noindent
In order to rigorously make sense of the integral appearing in the first term of Eq.~(\ref{general_propagator_mom}), one introduces the following notion of distributional convolution~\cite{Bogolubov_Logunov_Oksak90}:
\begin{align}
\left( \frac{1}{p^{2}+i\epsilon} \ast \rho, f \right) := \left( \rho,  \frac{1}{-p^{2}+i\epsilon} \ast f \right), 
\label{conv_def}
\end{align}
where $\frac{1}{p^{2}+i\epsilon} \ast \rho = \int ds \frac{\rho(s)}{p^{2}-s +i\epsilon}$, and $(D,f):= \int d^{4}x \, D(x)f(x)$ represents the smearing of the distribution $D$ with the test function $f$. For this definition to make sense for all test functions $f \in \mathcal{S}$, this requires that $\rho$ is extended from the class $\mathcal{S}'(\overbar{\mathbb{R}}_{+})$, as defined in Sec.~\ref{vect_corr_sect}, to the class $\mathcal{S}'(\overbar{\mathbb{R}}_{+} \cup \infty)$. In other words, the distribution $\rho$ must be permitted to have support at (positive) infinity. The origin of this requirement stems from the fact that propagators contain a product between theta distributions and ordinary correlators [see Eq.~(\ref{t_ordered_expl})], which is in general ill-defined. By extending the domain of validity of $\rho$, and thus making sense of the convolution $\frac{1}{p^{2}+i\epsilon} \ast \rho$, this is equivalent to defining this product~\cite{Bogolubov_Logunov_Oksak90}. A direct consequence of this extension is that the constant function $f\equiv 1$ is now a valid test function for the spectral density [since $1 \in \mathcal{S}(\overbar{\mathbb{R}}_{+} \cup \infty)$], and this therefore guarantees that the expressions $\int ds \rho_{2}(s)$ and $\int ds \rho_{1}(s)$ are both well defined. \\

\noindent
An important property of the representations in Eqs.~(\ref{general_propagator_pos}) and~(\ref{general_propagator_mom}) is that they follow only from the assumption that Fourier transformed correlators are Lorentz covariant tempered distributions with support in $\overbar{V}^{+}$. Since this is a ubiquitous feature of any axiomatically defined QFT, this means that these representations are model independent. Therefore, in order to further constrain the structure of particular propagators, one must introduce dynamical information about the fields $A_{\mu}$, such as equations of motion or (anti-)commutation relations. In Sec.~\ref{vector_examples} these constraints will be outlined in the cases where $A_{\mu}$ is a free photon field, the photon field in QED, and the gluon field in QCD, and the effect that they have on the form of the corresponding propagators will be discussed.

\section{Explicit vector propagators} 
\label{vector_examples}

\subsection{The free photon propagator}
\label{free_photon}

When $A_{\mu}$ is a free (locally) quantised electromagnetic field, it satisfies the following equations of motion:
\begin{align}
\partial^{\nu}F_{\nu\mu} + \partial_{\mu}\Lambda=0, \hspace{10mm} \xi\Lambda = \partial^{\mu}A_{\mu},
\end{align}  
where $\Lambda$ is a gauge fixing auxiliary field. As with any free theory, quantisation is performed by imposing equal-time commutation relations (ETCRs), which in this case are
\begin{align}
&\left[\Lambda(x),\Lambda(y)\right]_{x_{0}=y_{0}} = 0, \\
&\left[\Lambda(x),A_{\nu}(y)\right]_{x_{0}=y_{0}} = ig_{0\nu}\delta(\mathbf{x}-\mathbf{y}), \label{etcr1} \\
&\left[F_{0i}(x),A_{\nu}(y)\right]_{x_{0}=y_{0}} =  ig_{i\nu}\delta(\mathbf{x}-\mathbf{y}), \label{etcr2} \\
&\left[A_{\mu}(x),A_{\nu}(y)\right]_{x_{0}=y_{0}} = 0.  \label{etcr3}
\end{align}
It follows immediately from the equations of motion that: $\partial^{2}\Lambda = -\partial^{\mu}\partial^{\nu}F_{\nu\mu} = 0$, and thus $\Lambda$ satisfies a free wave equation. Among other things, this implies that any unequal-time commutator involving the field $\Lambda$ is uniquely determined (as a distribution) by the corresponding equal-time commutator~\cite{Strocchi13}. In particular, one has
\begin{align}
&\left[\Lambda(x),\Lambda(y)\right] = 0, \\
&\left[\Lambda(x),A_{\nu}(y)\right] = i\partial_{\nu}^{x}D_{0}(x-y). \label{ETCR_free}
\end{align}     
Moreover, since $\Lambda$ is a free field, one can decompose it into positive and negative frequency components: $\Lambda=\Lambda^{+}+\Lambda^{-}$, where the gauge fixing (subsidiary) condition corresponds to: $\Lambda^{-}\mathcal{V}_{\text{phys}}=0$. In order to constrain the form of the photon correlator, one can use the fact that the vacuum state is physical, from which it follows that
\begin{align}
&\langle 0| \Lambda(x)\Lambda(y)|0\rangle = 0, \label{constr1} \\
&\langle 0| \Lambda(x)A_{\nu}(y)|0\rangle = \langle 0| \left[\Lambda^{-}(x),A_{\nu}(y)\right]|0\rangle = i\partial_{\nu}^{x}D_{0}^{-}(x-y). \label{constr2}
\end{align}
\ \\
\noindent
Now that the equations of motion and ETCRs have been defined, one can establish the constraints that these relations impose on the structure of the free photon correlator and propagator. Firstly, using the equation of motion $\xi\Lambda = \partial^{\mu}A_{\mu}$, Eq.~(\ref{constr1}) can be written in the form 
\begin{align*}
\langle 0| \partial^{\mu}A_{\mu}(x)\partial^{\nu}A_{\nu}(y)|0\rangle= \partial^{\mu}_{x}\partial^{\nu}_{y}\langle 0| A_{\mu}(x)A_{\nu}(y)|0\rangle=0.
\end{align*} 
By inserting in Eq.~(\ref{vector_decomp_pos}), and taking the inverse Fourier transform of this expression, this then implies the equality 
\begin{align}
\theta(p^{0})p^{2} \left[ \rho_{1}(p^{2}) + p^{2} \rho_{2}(p^{2})\right] + \left[p^{2}\left(\sum_{l=0}^{L}\tilde{a}_{l}(\partial^{2})^{l}\right) + (p^{2})^{2}\left(\sum_{m=2}^{M}b_{m}(\partial^{2})^{m}\right)  \right]\delta(p) = 0.
\label{sum_constr}
\end{align}
\noindent
Since the first distribution in the equality above is defined to have support outside $p=0$ (in the closed forward light cone)~\cite{Bogolubov_Logunov_Oksak90}, whereas the second distribution has support at $p=0$, the equality requires that both distributions must vanish identically. It turns out that the vanishing of the first term in Eq.~(\ref{sum_constr}) implies the relation
\begin{align}
& \rho_{1}(s) + s \rho_{2}(s) = C\delta(s), \label{relation2}
\end{align} 
where $C$ is an arbitrary constant. Moreover, by using the distributional properties of $\delta(p)$ (and its derivatives), one can write
\begin{align}
&p^{2}\left(\sum_{l=0}^{L}\tilde{a}_{l}(\partial^{2})^{l}\right)\delta(p) = \sum_{l=1}^{L}4l(l+1)\tilde{a}_{l}(\partial^{2})^{l-1}\delta(p),  \\
&(p^{2})^{2}\left(\sum_{m=2}^{M}b_{m}(\partial^{2})^{m}\right)\delta(p) = \sum_{m=2}^{M}16m^{2}(m-1)(m+1)b_{m}(\partial^{2})^{m-2}\delta(p).
\end{align}
Setting $N:= \text{min}\{L-1,M-2 \}$ and $K:=\text{max}\{L-1,M-2 \}$, the vanishing of the second term then implies 
\begin{align}
&\tilde{a}_{n} = -4(n+1)(n+2)b_{n+1}, \hspace{5mm}  1 \leq n \leq N+1,  \label{relation1_1} \\
&\left.
 \begin{array}{ll}
       \tilde{a}_{n}=0, & \text{if \ $M < L+1$} \\
       b_{n+1}=0, &  \text{if \ $L+1 < M$}
     \end{array}
     \right\} \hspace{3mm} N+2 \leq n \leq K+1. \label{relation1_2} 
\end{align}
The constraint in Eq.~(\ref{constr1}) therefore ensures that the coefficients of the polynomials $\widetilde{P}_{i}$, as well as the spectral densities $\rho_{i}$, are no longer independent, but are in fact related to one another. \\

\noindent
The next constraint on the free photon correlator and propagator arises from Eq.~(\ref{constr2}). Again, by using the equation of motion $\xi\Lambda = \partial^{\mu}A_{\mu}$, this equation can be written
\begin{align*}
\partial^{\mu}_{x}\langle 0| A_{\mu}(x)A_{\nu}(y)|0\rangle &= \xi \langle 0| \Lambda(x)A_{\nu}(y)|0\rangle = i\xi\partial_{\nu}^{x}D_{0}^{-}(x-y).
\end{align*} 
Inserting Eq.~(\ref{vector_decomp_pos}), and then taking the inverse Fourier transform of this expression, implies the equality 
\begin{align*}
\theta(p^{0})p_{\nu} \left[\rho_{1}(p^{2}) + p^{2} \rho_{2}(p^{2}) + 2\pi\xi\delta(p^{2}) \right] + p_{\nu}\left[ \left(\sum_{l=0}^{L}\tilde{a}_{l}(\partial^{2})^{l}\right) + p^{2}\left( \sum_{m=2}^{M}b_{m}(\partial^{2})^{m} \right)   \right]\delta(p)  =  0.
\end{align*}
Just as with Eq.~(\ref{sum_constr}), both of the terms in this expression must vanish separately. Using the distributional identities:
\begin{align}
&p_{\nu}\left(\sum_{l=0}^{L}\tilde{a}_{l}(\partial^{2})^{l}\right)\delta(p) = \sum_{l=1}^{L}2l \, \tilde{a}_{l}\partial_{\nu}(\partial^{2})^{l-1}\delta(p),  \\
& p_{\nu}p^{2}\left(\sum_{m=2}^{M}b_{m}(\partial^{2})^{m}\right)\delta(p) =  \sum_{m=2}^{M}8m(m-1)(m+1)b_{m}\partial_{\nu}(\partial^{2})^{m-2}\delta(p),
\end{align}
it turns out that the vanishing of the second term implies identical constraints to those in Eqs.~(\ref{relation1_1}) and~(\ref{relation1_2}). Furthermore, by considering the $\nu=0$ component of the first term, and using the constraint in Eq.~(\ref{relation2}), one obtains
\begin{align*}
\theta(p^{0})p_{0} \left[\rho_{1}(p^{2}) + p^{2} \rho_{2}(p^{2}) + 2\pi\xi\delta(p^{2}) \right]   &=  \theta(p^{0})p_{0} \left[(C + 2\pi\xi)\delta(p^{2}) \right]  \\
 &=  p_{0} \left[(C + 2\pi\xi)\frac{\delta(p_{0}-|\mathbf{p}|)}{2|\mathbf{p}|} \right] = \frac{1}{2}(C + 2\pi\xi) = 0,
\end{align*}
and thus the constant in Eq.~(\ref{relation2}) is fixed to $C= -2\pi\xi$. In summary, the correlators in Eqs.~(\ref{constr1}) and~(\ref{constr2}) imply the following conditions:  
\begin{align}
&\tilde{a}_{n} = -4(n+1)(n+2)b_{n+1}, \hspace{5mm}  1 \leq n \leq N+1,  \label{final_relation1_1} \\
&\left.
 \begin{array}{ll}
       \tilde{a}_{n}=0, & \text{if \ $M < L+1$} \\
       b_{n+1}=0, &  \text{if \ $L+1 < M$}
     \end{array}
     \right\} \hspace{3mm} N+2 \leq n \leq K+1,  \label{final_relation1_2} \\
& \rho_{1}(s) + s \rho_{2}(s) = -2\pi\xi\delta(s). \label{final_relation2}
\end{align}
\ \\

\noindent
Although the constraints imposed by the relations in Eqs.~(\ref{constr1}) and~(\ref{constr2}) imply that the coefficients of the polynomials $\widetilde{P}_{1}$ and $\widetilde{P}_{2}$ are related to one another [Eqs.~(\ref{final_relation1_1}) and~(\ref{final_relation1_2})], these coefficients can still in principle be any complex numbers. However, it will now be demonstrated that further constraints on these parameters arise due to another important feature of free electromagnetism -- the field strength tensor $F_{\mu\nu}$ is an observable. The precise definition of operator observability is discussed in~\cite{Nakanishi_Ojima90}, but essentially because $F_{\mu\nu}$ is gauge invariant this is sufficient to imply it is an observable, and hence: $F_{\mu\nu}\mathcal{V}_{\text{phys}} \subseteq \mathcal{V}_{\text{phys}}$. Since by definition: $|\Psi\rangle \in \mathcal{V}_{\text{phys}} \Rightarrow \langle \Psi |\Psi\rangle \geq 0$, the observability of $F_{\mu\nu}$ and the fact that $|0\rangle \in \mathcal{V}_{\text{phys}}$ therefore gives rise to the following constraint:
\begin{align}
\langle 0|F(f)^{\dagger}F(f)|0\rangle \geq 0,
\label{f_constr}
\end{align}  
where $F(f):= \int d^{4}x F_{\mu\nu}(x)f^{\mu\nu}(x)$, with $f^{\mu\nu} \in \mathcal{S}(\mathbb{R}^{1,3})$. Because $F_{\mu\nu}=\partial_{\mu}A_{\nu}-\partial_{\nu}A_{\mu}$, one can write
\begin{align}
\langle 0|F_{\mu\nu}(x)F_{\rho\sigma}(y)|0 \rangle &=  \partial_{\mu}^{x}\partial_{\rho}^{y}\langle 0|A_{\nu}(x)A_{\sigma}(y)|0\rangle -\partial_{\mu}^{x}\partial_{\sigma}^{y}\langle 0|A_{\nu}(x)A_{\rho}(y)|0\rangle \nonumber \\
& \hspace{5mm} -\partial_{\nu}^{x}\partial_{\rho}^{y}\langle 0|A_{\mu}(x)A_{\sigma}(y)|0\rangle +\partial_{\nu}^{x}\partial_{\sigma}^{y}\langle 0|A_{\mu}(x)A_{\rho}(y)|0\rangle.
\label{ff_eqn}
\end{align} 
Moreover, due to Eq.~(\ref{vector_decomp_pos}) the vector correlator has the following form:
\begin{align}
\langle 0| A_{\mu}(x)A_{\nu}(y)|0\rangle  &= \partial_{\mu}^{x}\partial_{\nu}^{y}\underbrace{\left[ \frac{\widetilde{P}_{2}\left(-(x-y)^{2}\right)}{(2\pi)^{4}}-\int_{0}^{\infty} \frac{ds}{2\pi}  \rho_{2}(s)iD^{(-)}(x-y;s)   \right]}_{:= G(x-y)} \nonumber \\
& \hspace{10mm} +g_{\mu\nu}\underbrace{\left[ \frac{\widetilde{P}_{1}\left(-(x-y)^{2}\right)}{(2\pi)^{4}}-\int_{0}^{\infty} \frac{ds}{2\pi}  \rho_{1}(s) i D^{(-)}(x-y;s) \right]}_{:= F(x-y)},
\label{struct_FG}
\end{align}
which upon substitution into Eq.~(\ref{ff_eqn}) gives
\begin{align}
\langle 0|F_{\mu\nu}(x)F_{\rho\sigma}(y)|0 \rangle = \big( g_{\nu\sigma}\partial_{\mu}^{x}\partial_{\rho}^{y} -g_{\nu\rho}\partial_{\mu}^{x}\partial_{\sigma}^{y}  -g_{\mu\sigma}\partial_{\nu}^{x}\partial_{\rho}^{y} +g_{\mu\rho}\partial_{\nu}^{x}\partial_{\sigma}^{y} \big)F(x-y).
\label{g_F}
\end{align}
So the $G(x-y)$ component of the vector correlator does not contribute to the field strength correlator. Since $F(f)^{\dagger} = F(\bar{f})$, Eq.~(\ref{g_F}) can then be used to write the observability condition in Eq.~(\ref{f_constr}) as follows:
\begin{align}
\langle 0|F(f)^{\dagger}F(f)|0\rangle &=\int d^{4}xd^{4}y \ \langle 0|F_{\mu\nu}(x)F_{\rho\sigma}(y)|0 \rangle \, \bar{f}^{\mu\nu}(x)f^{\rho\sigma}(y) \nonumber \\
&= \int d^{4}xd^{4}y \ F(x-y)\bar{h}^{\rho}(x) h_{\rho}(y) \geq 0,
\label{pos_def}
\end{align} 
where $h_{\rho} := \partial_{\mu}f^{\mu}_{\ \rho} - \partial_{\mu}f_{\rho}^{\ \mu} \in \mathcal{S}(\mathbb{R}^{1,3})$. Since $h_{\rho}$ is an arbitrary test function, Eq.~(\ref{pos_def}) implies that $F(x-y)$ must be a \textit{positive-definite} distribution. An important feature of positive-definitive distributions is that their Fourier transform $\widehat{F}(p)$ is a \textit{non-negative} distribution, and this in turn defines a measure~\cite{Bogolubov_Logunov_Oksak90}. Since $\widehat{F}(p) = \int_{0}^{\infty} ds \, \rho_{1}(s)\theta(p^{0})\delta(p^{2}-s) + \widetilde{P}_{1}(\partial^{2})\delta(p)$, in particular this means that $\widetilde{P}_{1}(\partial^{2})\delta(p)$ \textit{cannot} contain terms involving derivatives of $\delta(p)$, because these distributions do not define measures~\cite{Baake_Grimm13}, and thus one must have: $\tilde{a}_{k}=0$ $\forall k\geq 1$. Taken together with the relations in Eqs.~(\ref{final_relation1_1}) and~(\ref{final_relation1_2}), this therefore implies the following constraint on the polynomial coefficients:
\begin{align}
\tilde{a}_{k} = b_{k+1} = 0, \hspace{5mm} \forall k \geq 1.
\label{ab_constr}
\end{align}     
Due to the definitions of the polynomial terms in Eq.~(\ref{vector_decomp_pos}), an immediate corollary of this constraint is that: $\widetilde{P}_{2}=0$, and $\widetilde{P}_{1} = \tilde{a}_{0}$. In other words, the polynomial terms only contribute to the free photon correlator or propagator if $\tilde{a}_{0}$  is non-vanishing. \\

\noindent
In principle the coefficient $\tilde{a}_{0}$ could be non-vanishing, but it turns out that $\widehat{F}(p)$ defining a measure guarantees that this is not the case. To see this, consider the following (cluster) correlator: 
\begin{align*}
\langle 0|\widetilde{F}(\tilde{x})\widetilde{F}(\tilde{y})|0 \rangle := \int d^{4}xd^{4}y \ \langle 0|F_{\mu\nu}(x)F_{\rho\sigma}(y)|0 \rangle \, \bar{f}^{\mu\nu}(x-\tilde{x})f^{\rho\sigma}(y-\tilde{y}).
\end{align*}  
\noindent 
Taking the Fourier transform of this expression, and applying Eq.~(\ref{g_F}), gives
\begin{align*}
\mathcal{F}\left[\langle 0|\widetilde{F}(\tilde{x})\widetilde{F}(\tilde{y})|0 \rangle\right]  =\mathcal{F}\left[\int d^{4}xd^{4}y \ F(x-y)\bar{h}^{\rho}(x-\tilde{x}) h_{\rho}(y-\tilde{y})\right]  = \hat{\bar{h}}^{\rho}(-p) \hat{h}_{\rho}(p)\widehat{F}(p).
\end{align*}  
Since $\widehat{F}(p)$ defines a measure it follows that $\mathcal{F}\left[\langle 0|\widetilde{F}(\tilde{x})\widetilde{F}(\tilde{y})|0 \rangle\right]$ must \textit{also} define a measure~\cite{Lowdon16}. Moreover, due to Eq.~(\ref{ab_constr}), this measure has the contribution $\tilde{a}_{0}\hat{\bar{h}}^{\rho}(0) \hat{h}_{\rho}(0)\delta(p)$ at the point $p=0$. However, one of the Pseudo-Wightman axioms~\cite{Bogolubov_Logunov_Oksak90} states that since the Fourier transform of $\langle 0|\widetilde{F}(\tilde{x})\widetilde{F}(\tilde{y})|0 \rangle$ defines a (complex) measure, it must be the case that the contribution of this measure at the point $p=0$ is equal to $(2\pi)^{4} \langle 0|\widetilde{F}(\tilde{x})|0\rangle \langle 0|\widetilde{F}(\tilde{y})|0 \rangle \delta(p)$. Therefore, one must have the equality
\begin{align*}
\tilde{a}_{0}\hat{\bar{h}}^{\rho}(0) \hat{h}_{\rho}(0) &= (2\pi)^{4} \langle 0|\widetilde{F}(\tilde{x})|0\rangle \langle 0|\widetilde{F}(\tilde{y})|0 \rangle \\
&= (2\pi)^{4} \int d^{4}xd^{4}y \ \langle 0|F_{\mu\nu}(x)|0\rangle \langle 0|F_{\rho\sigma}(y)|0 \rangle  \bar{f}^{\mu\nu}(x-\tilde{x})f^{\rho\sigma}(y-\tilde{y}).
\end{align*}
But $\langle 0|F_{\mu\nu}(x)|0\rangle=\langle 0|F_{\rho\sigma}(y)|0 \rangle =0$ because one cannot have a non-Lorentz invariant condensate, and so it must be that: $\tilde{a}_{0}=0$. Combining this constraint with Eq.~(\ref{ab_constr}) implies:
\begin{align}
\widetilde{P}_{1}=\widetilde{P}_{2}=0.
\label{p_constr}
\end{align}     

\ \\

\noindent
Another constraint on the form of the vector correlator, and in particular the spectral densities $\rho_{i}$, arises from the equal-time commutation relation
\begin{align}
\left[A_{\mu}(x),\dot{A}_{\nu}(y)\right]_{x_{0}=y_{0}} = -i\left[g_{\mu\nu} - (1-\xi)g_{0\mu}g_{0\nu} \right]\delta(\mathbf{x}-\mathbf{y}),
\label{etcr_aa} 
\end{align}
which itself is derived from the equations of motion and Eqs.~(\ref{etcr1}),~(\ref{etcr2}) and~(\ref{etcr3}). Setting $\mu=i,\nu =j$ one has that
\begin{align*}
\left[\partial_{y}^{0}\langle 0|A_{i}(x)A_{j}(y)|0\rangle - \partial_{y}^{0} \langle 0|A_{j}(y)A_{i}(x)|0\rangle \right]_{x_{0}=y_{0}}  = -ig_{ij}\delta(\mathbf{x}-\mathbf{y}).
\end{align*} 
Inserting in the general expression for the correlator in Eq.~(\ref{vector_decomp_pos}), one obtains the following sum rules:
\begin{align}
\int_{0}^{\infty} ds  \, \rho_{1}(s)  = -2\pi, \hspace{5mm} \int_{0}^{\infty} ds  \, \rho_{2}(s)  = 0. 
\label{sum_rules_rho_1}
\end{align}
One should note here that even if the polynomial terms $\widetilde{P}_{i}$ were non-vanishing, they would cancel in the commutator and hence not affect the constraints in Eq.~(\ref{sum_rules_rho_1}). Similarly, in the case where $\mu=\nu=0$, this instead implies the sum rules
\begin{align}
\int_{0}^{\infty} ds  \, \left[\rho_{1}(s) + s\rho_{2}(s)\right]  = -2\pi\xi, \hspace{5mm} \int_{0}^{\infty} ds  \, \rho_{2}(s)  = 0.
\label{sum_rules_rho_2}
\end{align}
So both the constraints imply that the integral of the spectral density $\rho_{2}$ vanishes, whereas Eq.~(\ref{sum_rules_rho_1}) constrains the integral of $\rho_{1}$, and Eq.~(\ref{sum_rules_rho_2}) constrains the integral of the combination $\rho_{1} + s\rho_{2}$. \\     

\noindent
A final constraint on the form of the free photon correlator arises because the equation of motion can be written: $\partial^{\nu}F_{\nu\mu} + \partial_{\mu}\Lambda= \partial^{2}A_{\mu} + (1-\xi)\partial_{\mu}\Lambda = 0$, which means that 
\begin{align*}
\partial^{2}\langle 0| A_{\mu}(x)A_{\nu}(y)|0\rangle &= (\xi-1)\partial_{\mu}^{x}\langle 0| \Lambda(x)A_{\nu}(y)|0\rangle = i(\xi-1)\partial_{\mu}^{x}\partial_{\nu}^{x}D_{0}^{-}(x-y).
\end{align*}
By inserting the general expression for the correlator in Eq.~(\ref{vector_decomp_pos}), as well as the constraint $\widetilde{P}_{1}=\widetilde{P}_{2}=0$, and taking the inverse Fourier transform, this equality implies
\begin{align*}
\theta(p^{0})\big[ g_{\mu\nu}p^{2}\rho_{1}(p^{2}) + p_{\mu}p_{\nu}p^{2}\rho_{2}(p^{2})  + 2\pi(\xi-1)p_{\mu}p_{\nu}\delta(p^{2}) \big]   =0.
\end{align*}
Substituting in the condition on the spectral densities in Eq.~(\ref{final_relation2}) into this relation, one obtains
\begin{align*}
\theta(p^{0})\left[ (g_{\mu\nu}p^{2} - p_{\mu}p_{\nu})\rho_{1}(p^{2}) -2\pi p_{\mu}p_{\nu}\delta(p^{2}) \right] =0,
\end{align*} 
which upon contraction with $g^{\mu\nu}$ implies
\begin{align*}
\theta(p^{0})\left[3p^{2}\rho_{1}(p^{2}) - 2\pi p^{2}\delta(p^{2})\right]  = 3\theta(p^{0})p^{2}\rho_{1}(p^{2}) = 0,
\end{align*}  
and hence: $\rho_{1}(p^{2}) = D\delta(p^{2})$ for some arbitrary constant $D$. By applying the sum rule for $\rho_{1}$ in Eq.~(\ref{sum_rules_rho_1}) it immediately follows that $D = -2\pi$. Since $\rho_{1}(p^{2}) = -2\pi\delta(p^{2})$, this means that $\rho_{2}$ satisfies the equation 
\begin{align}
p^{2}\rho_{2}(p^{2})= 2\pi(1 - \xi)\delta(p^{2}). 
\end{align}
The general solution to this equation has the form: $\rho_{2}(p^{2}) = E\delta(p^{2}) -2\pi(1 - \xi)\delta'(p^{2})$, where $E$ is an arbitrary constant. It follows from the sum for $\rho_{2}$ in Eq.~(\ref{sum_rules_rho_1}) that $E=0$ and thus one can finally conclude that the spectral densities for the free photon correlator have the following exact form:
\begin{align}
\rho_{1}(s) = -2\pi\delta(s), \hspace{5mm} \rho_{2}(s)= -2\pi(1- \xi)\delta'(s). 
\label{rhos_free_photon}
\end{align}
Given these spectral densities, and the fact that $\widetilde{P}_{1}=\widetilde{P}_{2}=0$, the momentum space free photon correlator can therefore be written
\begin{align}
\widehat{D}_{\mu\nu}(p) = 2\pi\theta(p^{0}) \left[ -g_{\mu\nu}\delta(p^{2}) + p_{\mu}p_{\nu}(\xi-1)\delta'(p^{2}) \right]. 
\label{free_photon_corr}
\end{align}
Moreover, since the constraints from Eq.~(\ref{etcr_aa}) imply that the integral of $\rho_{2}$ vanishes, it follows from Eq.~(\ref{general_propagator_mom}) that the free photon propagator has the form
\begin{align}
\widehat{D}_{\mu\nu}^{F}(p) &=   i\int_{0}^{\infty} \frac{ds}{2\pi} \, \frac{\left[ g_{\mu\nu}\rho_{1}(s) + p_{\mu}p_{\nu}\rho_{2}(s) \right]}{p^{2}-s +i\epsilon}, 
\end{align}
which upon substitution of the expressions for $\rho_{1}$ and $\rho_{2}$ in Eq.~(\ref{rhos_free_photon}) gives
\begin{align}
\widehat{D}_{\mu\nu}^{F}(p) &= -\left[g_{\mu\nu} - (1-\xi)\frac{p_{\mu}p_{\nu}}{p^{2}+i\epsilon} \right] \frac{i}{p^{2}+i\epsilon} \nonumber \\
&= -\underbrace{\left(g_{\mu\nu} - \frac{p_{\mu}p_{\nu}}{p^{2}+i\epsilon} \right)}_{:=T_{\mu\nu}} \frac{i}{p^{2}+i\epsilon}   -i\xi \underbrace{\frac{p_{\mu}p_{\nu}}{(p^{2}+i\epsilon)^{2}}}_{:=L_{\mu\nu}},
\label{free_photon_prop}
\end{align}  
where $T_{\mu\nu}$ and $L_{\mu\nu}$ are referred to as the \textit{transverse} and \textit{longitudinal} projectors respectively. \\

\subsection{The photon propagator in QED}
\label{QED_photon}

In QED one requires the fields to be renormalised in order to make sense of the equations of motion. Once this renormalisation has been performed, the equations of motion in locally quantised QED have the following form: 
\begin{align}
\partial^{\nu}F_{\nu\mu}^{(r)} + \partial_{\mu}\Lambda^{(r)} =j_{\mu}^{(r)}, \hspace{5mm} \xi_{r}\Lambda^{(r)} = \partial^{\mu}A_{\mu}^{(r)},
\end{align} 
where the index $r$ indicates that the corresponding quantity is renormalised, and $j_{\mu}^{(r)}$ is the (conserved) fermion interaction current. In particular, one has that $A_{\mu}^{(r)}=Z_{3}^{-\frac{1}{2}}A_{\mu}^{(0)}$, where $Z_{3}$ is the photon field renormalisation constant and $A_{\mu}^{(0)}$ is the unrenormalised bare field. For simplicity, throughout the rest of this paper the label $r$ will be dropped, and every quantity should be implicitly assumed to be renormalised. To quantise QED one imposes the ETCRs:
\begin{align}
&\left[\Lambda(x),\Lambda(y)\right]_{x_{0}=y_{0}} = 0, \\
&\left[\Lambda(x),A_{\nu}(y)\right]_{x_{0}=y_{0}} = ig_{0\nu}\delta(\mathbf{x}-\mathbf{y}), \\
&\left[F_{0i}(x),A_{\nu}(y)\right]_{x_{0}=y_{0}} =  ig_{i\nu}Z_{3}^{-1}\delta(\mathbf{x}-\mathbf{y}), \\
&\left[A_{\mu}(x),A_{\nu}(y)\right]_{x_{0}=y_{0}} = 0. 
\end{align} 
An important feature here is that even though the equation of motion includes the non-vanishing current $j_{\mu}$, $\Lambda$ still satisfies the free massless wave equation by virtue of the current conservation condition $\partial^{\mu}j_{\mu}=0$. Among other things, this implies that the renormalisation constant $Z_{3}$ must be finite~\cite{Strocchi13}, and therefore the correlators involving the auxiliary field $\Lambda$ are the same as those in the free case (Eqs.~(\ref{constr1}) and~(\ref{constr2}): 
\begin{align}
&\langle 0| \Lambda(x)\Lambda(y)|0\rangle = 0, \label{qed_constr1} \\
&\langle 0|\Lambda(x)A_{\nu}(y)|0\rangle =  i\partial_{\nu}^{x}D_{0}^{-}(x-y). \label{qed_constr2}
\end{align} 
Moreover, because $F_{\mu\nu}$ is gauge invariant, it follows that $F_{\mu\nu}$ is \textit{also} an observable in QED. Since the structural relations for vector correlators and propagators derived in Sec.~\ref{np_gen} are equally applicable to both free and interacting theories, the constraints implied by the observability of $F_{\mu\nu}$ and Eqs.~(\ref{qed_constr1}) and~(\ref{qed_constr2}) are identical to those in the free photon case:  
\begin{align}
&\widetilde{P}_{1}=\widetilde{P}_{2}=0, \\
&\rho_{1}(s) + s \rho_{2}(s) = -2\pi \xi\delta(s), \label{photon_rho_rel} \\
\int_{0}^{\infty} ds  \, \rho_{1}(s)  = -2\pi Z_{3}^{-1},  \hspace{2mm} &\int_{0}^{\infty} ds  \, \left[\rho_{1}(s) + s\rho_{2}(s)\right]  = -2\pi\xi, \hspace{2mm} \int_{0}^{\infty} ds  \, \rho_{2}(s)  = 0.  \label{sum_rules_photon_rho}
\end{align}

\noindent
Using the above constraints, it follows analogously to Sec.~\ref{free_photon} that the momentum space photon correlator has the structure:
\begin{align}
\widehat{D}_{\mu\nu}(p) =  \int_{0}^{\infty} ds \, \theta(p^{0})\delta(p^{2}-s) \left[ g_{\mu\nu}\rho_{1}(s) + p_{\mu}p_{\nu} \rho_{2}(s) \right],
\label{photon_corr}
\end{align}  
and hence the photon propagator can be written 
\begin{align}
\widehat{D}_{\mu\nu}^{F}(p) &=   i\int_{0}^{\infty} \frac{ds}{2\pi} \, \frac{\left[ g_{\mu\nu}\rho_{1}(s) + p_{\mu}p_{\nu}\rho_{2}(s) \right]}{p^{2}-s +i\epsilon}. 
\label{general_propagator_QED_mom}
\end{align}
An important feature of the spectral densities in QED, as opposed to the free case, is that despite being related to one another via Eq.~(\ref{photon_rho_rel}), the explicit form of the spectral densities is \textit{not} determined. This lack of knowledge arises because of the non-trivial non-perturbative structure of the theory.

\subsection{The gluon propagator in QCD}
\label{QCD_prop}

In BRST quantised QCD, the equations of motion have the following form:
\begin{align}
&(D^{\nu}F_{\nu\mu})^{a} +\partial_{\mu}\Lambda^{a} = gj_{\mu}^{a} -igf^{abc}\partial_{\mu}\overbar{C}^{b}C^{c}, \hspace{3mm} \partial^{\mu}A_{\mu}^{a} = \xi\Lambda^{a}, \\
&\partial^{\nu}(D_{\nu}C)^{a}=0, \hspace{5mm} (D^{\nu}\partial_{\nu}\overbar{C})^{a}=0,
\end{align}
where $C^{a}$ and $\overbar{C}^{a}$ are the ghost and anti-ghost fields, and all of the fields depend on the non-abelian adjoint index $a$. The ETCRs of particular relevance are
\begin{align}
&\left[\Lambda^{a}(x),\Lambda^{b}(y)\right]_{x_{0}=y_{0}} = 0, \\
&\left[\Lambda^{a}(x),A_{\nu}^{b}(y)\right]_{x_{0}=y_{0}} = i\delta^{ab}g_{0\nu}\delta(\mathbf{x}-\mathbf{y}), \\
&\left[F_{0i}^{a}(x),A_{\nu}^{b}(y)\right]_{x_{0}=y_{0}} =  i\delta^{ab}g_{i\nu}Z_{3}^{-1}\delta(\mathbf{x}-\mathbf{y}), \\
&\left[A_{\mu}^{a}(x),A_{\nu}^{b}(y)\right]_{x_{0}=y_{0}} = 0,
\end{align}  
where now $Z_{3}$ is the gluon field renormalisation constant. Although these ETCRs have a similar form to those in QED and the free case, there is a very important difference in QCD -- the auxiliary field $\Lambda^{a}$ \textit{does not} satisfy a free wave equation. This means that unlike in QED and free electromagnetism, the ETCRs involving the auxiliary field \textit{cannot} be used to determine the value of the commutators at unequal times. In particular, one cannot assume that Eq.~(\ref{ETCR_free}) holds. Nevertheless, one can use the BRST symmetry of the QCD equations of motion to prove that the auxiliary field correlator $\langle 0|\Lambda^{a}(x)\Lambda^{b}(y)|0\rangle$ does in fact vanish, just like in Secs.~\ref{free_photon} and~\ref{QED_photon}. The key to this derivation is that the BRST variation of any product of fields $\mathcal{O}$ vanishes
\begin{align*}
\langle 0| \delta_{B}\mathcal{O} |0\rangle=\langle 0| \left[iQ_{B},\mathcal{O} \right]_{\pm} |0\rangle=0.
\end{align*} 
This automatically follows from the fact that $Q_{B}|0\rangle=0$ since $|0\rangle \in \mathcal{V}_{\text{phys}}$. By taking $\mathcal{O}=\partial_{\mu}A^{\mu,a}(x)\overbar{C}^{b}(y)$ one has:
\begin{align*}
0 = \langle 0| \delta_{B}\left(\partial_{\mu}A^{\mu,a}(x)\overbar{C}^{b}(y)\right) |0\rangle &=\langle 0| \delta_{B}\left(\partial_{\mu}A^{\mu,a}(x)\right)\overbar{C}^{b}(y) |0\rangle  + \langle 0| \partial_{\mu}A^{\mu,a}(x)\delta_{B}\left(\overbar{C}^{b}(y)\right) |0\rangle \\
&= \langle 0|\partial_{\mu} \delta_{B}\left(A^{\mu,a}(x)\right)\overbar{C}^{b}(y) |0\rangle  + \langle 0| \partial_{\mu}A^{\mu,a}(x)\delta_{B}\left(\overbar{C}^{b}(y)\right) |0\rangle \\
&= \langle 0|\underbrace{\partial_{\mu}(D^{\mu}C(x))^{a}}_{=0}\overbar{C}^{b}(y) |0\rangle + \langle 0| \partial_{\mu}A^{\mu,a}(x)(-i\Lambda^{b}(y))|0\rangle \\
&= -i\langle 0| \partial_{\mu}A^{\mu,a}(x)\Lambda^{b}(y)|0\rangle.
\end{align*}      
Using the equation of motion: $\partial^{\mu}A_{\mu}^{a} = \xi\Lambda^{a}$ this then leads immediately to: $\langle 0| \Lambda^{a}(x)\Lambda^{b}(y)|0\rangle=0$. Just as in the case of QED, one can apply the same analysis as for free photon correlator and propagator in Sec.~\ref{free_photon}, and this leads to the analogous constraints
\begin{align}
&\tilde{a}_{n}^{ab} = -4(n+1)(n+2)b_{n+1}^{ab}, \hspace{5mm}  1 \leq n \leq N+1,   \label{ab_coeff_rel1} \\
&\left.
 \begin{array}{ll}
       \tilde{a}_{n}^{ab}=0, & \text{if \ $M < L+1$} \\
       b_{n+1}^{ab}=0, &  \text{if \ $L+1 < M$}
     \end{array}
     \right\} \hspace{3mm} N+2 \leq n \leq K+1,  \label{ab_coeff_rel2} \\
&\rho_{1}^{ab}(s) + s \rho_{2}^{ab}(s) = C^{ab}\delta(s), \label{gluon_rho_rel}
\end{align}
where now the spectral densities and coefficients of the polynomials $\widetilde{P}_{1}^{ab}$ and $\widetilde{P}_{2}^{ab}$ must depend explicitly on the adjoint indices $a$ and $b$, and one assumes that the colour symmetry is unbroken, and thus: $\rho_{i}^{ab} = \delta^{ab}\rho_{i}$. Although one does not have an expression like Eq.~(\ref{ETCR_free}) to determine the value of $C^{ab}$, as in the free case and QED, the ETCRs still give rise to the sum rules
\begin{align}
\int_{0}^{\infty} ds  \, \rho_{1}^{ab}(s)  = -2\pi\delta^{ab} Z_{3}^{-1}, \hspace{1mm} \int_{0}^{\infty} ds  \, \left[\rho_{1}^{ab}(s) + s\rho_{2}^{ab}(s)\right]  = -2\pi\xi\delta^{ab}, \hspace{1mm} \int_{0}^{\infty} ds  \, \rho_{2}^{ab}(s)  = 0, 
\label{sum_rules_gluon_rho}
\end{align}  
the second of which implies that $C^{ab} = -2\pi\xi\delta^{ab}$, and hence:
\begin{align}
\rho_{1}^{ab}(s) + s \rho_{2}^{ab}(s) = -2\pi\xi\delta^{ab}\delta(s).
\label{spectr_rel_gluon}
\end{align}
\ \\
\noindent
An important difference between QCD and QED (or the free case), is that $F_{\mu\nu}^{a}$ is no longer an observable. This means that although one can decompose the gluon correlator in an analogous manner to Eq.~(\ref{struct_FG})
\begin{align}
\langle 0| A_{\mu}^{a}(x)A_{\nu}^{b}(y)|0\rangle  &=  g_{\mu\nu}F^{ab}(x-y) + \partial_{\mu}^{x}\partial_{\nu}^{y}G^{ab}(x-y),
\label{struct_FG_qcd}
\end{align}  
one is not guaranteed that the Fourier transform of $F^{ab}(x-y)$ defines a measure. Since this property is essential for demonstrating that the coefficients of the polynomials $\widetilde{P}^{ab}_{i}$ vanish, as discussed in Sec.~\ref{free_photon}, it is therefore possible that these coefficients are related [via Eqs.~(\ref{ab_coeff_rel1}) and~\ref{ab_coeff_rel2})] but non-zero. In other words, the fact that $F^{ab}(x-y)$ does not necessarily define a measure implies that the polynomials $\widetilde{P}_{i}^{ab}$ can be non-vanishing, and hence the propagator is permitted to contain terms involving derivatives of $\delta(p)$. \\

\noindent
Due to the various constraints in Eqs.~(\ref{ab_coeff_rel1}),~(\ref{ab_coeff_rel2}) and~(\ref{sum_rules_gluon_rho}), it follows that the gluon propagator can be written in the following general form:
\begin{align}
\widehat{D}_{\mu\nu}^{ab\, F}(p) =   i\int_{0}^{\infty} \frac{ds}{2\pi} \, \frac{\left[ g_{\mu\nu}\rho_{1}^{ab}(s) + p_{\mu}p_{\nu}\rho_{2}^{ab}(s) \right]}{p^{2}-s +i\epsilon}  +\sum_{n=0}^{N+1} \left[ c_{n}^{ab} \, g_{\mu\nu} (\partial^{2})^{n} + d_{n}^{ab} \partial_{\mu}\partial_{\nu}(\partial^{2})^{n-1}\right]\delta(p),
\label{general_propagator_QCD_mom}
\end{align}
where the (complex) coefficients $c_{n}$ and $d_{n}$ are defined by:
\begin{align}
c_{n}^{ab} =  &\left\{
 \begin{array}{ll}
       -2(n+1)(2n+3)b_{n+1}^{ab}, & 1 \leq n \leq N+1 \\
       \tilde{a}_{0}^{ab}, &  n=0
     \end{array}
     \right. \\
d_{n}^{ab} =  &\left\{
 \begin{array}{ll}
       4n(n+1)b_{n+1}^{ab}, & \hspace{13mm} 1 \leq n \leq N+1 \\
       0, &  \hspace{13mm} n=0
     \end{array}
     \right. 
\end{align}   

\noindent
By contrast to the photon propagator, the gluon propagator is only specified up to $N+2$ arbitrary complex coefficients. In this case the dynamical constraints are \textit{not} sufficient to determine whether these coefficients are vanishing or not, and this ultimately stems from the fact $F_{\mu\nu}^{a}$ is no longer an observable in QCD. This therefore opens up the possibility that the gluon propagator can contain singular terms involving derivatives of $\delta(p)$. Derivatives of $\delta(p)$ have the property of not defining measures, unlike $\delta(p)$, and it turns out that this property allows the correlation strength between clusters of fields to \textit{increase} with separation~\cite{Lowdon16}. This mechanism is particularly interesting in the context of QCD, since a growth of the correlation strength (with increasing distance) between coloured particle-creating fields would be a sufficient condition for confinement. Therefore, the fact that $F_{\mu\nu}^{a}$ fails to define an observable, and hence permits derivative of $\delta(p)$ terms to exist, is suggestive that the non-abelian nature of the gauge symmetry may well play an important role in ensuring that confinement occurs in non-perturbative QCD. \\

\noindent
In the literature, the analysis of the gluon propagator is performed using a variety of different non-perturbative techniques, including the Schwinger-Dyson equation~\cite{Alkofer_vonSmekal01,Alkofer_Detmold_Fischer_Maris04,Strauss_Fischer_Kellermann12,Kizilersu_Sizer_Pennington_Williams_Williams15}, and lattice QCD~\cite{Cucchieri_Mendes_Taurines05,Cucchieri_Mendes08,Dudal_Oliveira_Silva14}. Since both of these approaches aim to uncover the characteristics of the propagator, it is important to understand whether the singular terms discussed previously can in fact be detected using these methods. In the case of the Schwinger-Dyson equation, the equation itself involves various terms, including products of the gluon propagator $\widehat{D}_{\mu\nu}^{ab\, F}(p)$ with various vertex terms $\Gamma(p)$. By introducing general ans\"{a}tze for $\Gamma(p)$, this enables the equation to be solved recursively. However, since the precise structure of these vertex terms is unknown, one cannot guarantee that the product $\Gamma(p)\widehat{D}_{\mu\nu}^{ab\, F}(p)$ is meaningful, particularly if $\widehat{D}_{\mu\nu}^{ab\, F}(p)$ contains singular terms. This issue arises because both $\Gamma(p)$ and $\widehat{D}_{\mu\nu}^{ab\, F}(p)$ are distributions, not functions, and so their product is not necessarily well-defined~\cite{Bogolubov_Logunov_Oksak90}. In order to illustrate this point, consider the situation where both of these objects contains a $\delta(p)$ contribution. The Schwinger-Dyson equation would then necessarily contain the ill-defined expression $\delta(p)\delta(p)$. In light of these possible ambiguities, it may well be the case that in order for the Schwinger-Dyson equation to possess a well-defined solution one must intrinsically assume that no such singular terms are present. Whether or not this casts doubt on the existence of these singular terms in the gluon propagator remains to be seen, but it certainly suggests that if these singular terms are indeed present, then this method would potentially have difficulties detecting them. In the case of lattice QCD, it also unclear as to whether singular distributional terms like $\delta(p)$ can be observed. Nevertheless, one can in principle probe quantities like the Schwinger function~\cite{Lowdon16}, which are \textit{indirectly} sensitive to the distributional behaviour of the propagator\footnote{It turns out that if the gluon propagator did indeed contain derivatives of $\delta(p)$, then these terms would introduce a polynomial $t^{2}$ dependence in the Schwinger function $C(t)$. See~\cite{Lowdon16} for more details about the definition of $C(t)$.}.

\section{The transverse-longitudinal decomposition of the photon and gluon propagators}
\label{tran_long}

In the literature, the structure of the photon and gluon propagators are often derived using the following \textit{Slavnov-Taylor} identity\footnote{In the case of QED this relation is referred to as the \textit{Ward-Takahashi} identity, and the adjoint indices $a,b$ are dropped (i.e. $\delta^{ab}=1$) because the gauge group is abelian.}:
\begin{align}
p^{\mu}p^{\nu}\widehat{D}_{\mu\nu}^{ab\, F}(p) = -i\xi\delta^{ab}.
\label{WT}
\end{align}
It is often claimed~\cite{Muta87,Gogokhia_Barnafoldi13} that Eq.~(\ref{WT}) implies that the photon and gluon propagators have the following general transverse-longitudinal structure:
\begin{align}
\widehat{D}_{\mu\nu}^{ab\, F}(p) = T_{\mu\nu}D^{ab}(p^{2}) -i\xi\delta^{ab} L_{\mu\nu} = \left(g_{\mu\nu} - \frac{p_{\mu}p_{\nu}}{p^{2}+i\epsilon} \right)D^{ab}(p^{2}) -i\xi\delta^{ab} \frac{p_{\mu}p_{\nu}}{(p^{2}+i\epsilon)^{2}},
\label{prop_can}
\end{align} 
where $D^{ab}(p^{2})$ is Lorentz invariant. In the case of the free photon propagator [Eq.~(\ref{free_photon_prop})] this structure is indeed present. However, for QED and QCD it will be argued in the proceeding section that the propagators \textit{cannot} in general be written in this form. \\

\noindent
The constraints imposed by the equations of motion and the ETCRs in QED and QCD imply that the photon and gluon propagators have the form of Eqs.~(\ref{general_propagator_QED_mom}) and~(\ref{general_propagator_QCD_mom}) respectively. As well as defining the general structure, the constraints on the photon and gluon propagators also imply that the spectral densities are related to one another [via Eqs.~(\ref{photon_rho_rel}) and~(\ref{spectr_rel_gluon})]. Therefore, one can attempt to write the photon and gluon propagators exclusively in terms of either $\rho_{1}^{ab}$ or $\rho_{2}^{ab}$. In terms of $\rho_{2}^{ab}$, the photon and gluon propagators have the form 
\begin{align}
&\widehat{D}_{\mu\nu}(p) =  i\int_{0}^{\infty} \frac{ds}{2\pi} \, \left( -sg_{\mu\nu} + p_{\mu}p_{\nu} \right)\frac{\rho_{2}(s)}{p^{2}-s +i\epsilon}  -\frac{ig_{\mu\nu}\xi}{p^{2}+i\epsilon},  \label{photon_rho2} \\
&\widehat{D}_{\mu\nu}^{ab\, F}(p) =  i\int_{0}^{\infty} \frac{ds}{2\pi} \, \left( -sg_{\mu\nu} + p_{\mu}p_{\nu} \right)\frac{\rho_{2}^{ab}(s)}{p^{2}-s +i\epsilon}  -\frac{ig_{\mu\nu}\xi\delta^{ab}}{p^{2}+i\epsilon} \nonumber \\
&\hspace{20mm} +\sum_{n=0}^{N+1} \left[ c_{n}^{ab} \, g_{\mu\nu} (\partial^{2})^{n} + d_{n}^{ab} \partial_{\mu}\partial_{\nu}(\partial^{2})^{n-1}\right]\delta(p). \label{gluon_rho2}
\end{align}
Contracting both of these representations with $p^{\mu}p^{\nu}$ one obtains
\begin{align*}
&p^{\mu}p^{\nu}\widehat{D}_{\mu\nu}(p)  = ip^{2}\int_{0}^{\infty} \frac{ds}{2\pi} \, \rho_{2}(s)  - i\xi =  -i\xi, \\
&p^{\mu}p^{\nu}\widehat{D}_{\mu\nu}^{ab\, F}(p)  = ip^{2}\int_{0}^{\infty} \frac{ds}{2\pi} \, \rho_{2}^{ab}(s)  - i\xi\delta^{ab} \\
& \hspace{30mm} + \underbrace{p^{\mu}p^{\nu}\sum_{n=0}^{N+1} \left[ c_{n}^{ab} \, g_{\mu\nu} (\partial^{2})^{n} + d_{n}^{ab} \partial_{\mu}\partial_{\nu}(\partial^{2})^{n-1}\right]\delta(p)}_{=0}  = -i\xi\delta^{ab},
\end{align*} 
where the last equality holds in both cases due to the $\rho_{2}^{ab}$ integral constraint in Eqs.~(\ref{sum_rules_photon_rho}) and~(\ref{sum_rules_gluon_rho}) respectively. This demonstrates that both the photon and gluon propagators do indeed satisfy Eq.~(\ref{WT}). Nevertheless, it is clear that both propagator representations in Eqs.~(\ref{photon_rho2}) and~(\ref{gluon_rho2}) do not have the form of Eq.~(\ref{prop_can}). The only other possibility to express these propagators in this form is to write them exclusively in terms of the spectral density $\rho_{1}^{ab}$. Since $\rho_{2}^{ab}$ and $\rho_{1}^{ab}$ are related by Eqs.~(\ref{photon_rho_rel}) and~(\ref{spectr_rel_gluon}), this problem boils down to solving the (distributional) equation
\begin{align}
s \rho_{2}^{ab}(s) = -2\pi \xi\delta^{ab}\delta(s) - \rho_{1}^{ab}(s). 
\label{constrain}
\end{align}
It turns out that this equation always possesses solutions~\cite{Hormander58}. In particular, one can write:
\begin{align*}
\int ds \, \rho_{2}^{ab}(s)f(s) &:= (\rho_{2}^{ab},f) = \mathcal{C}^{ab} f(0) -2\pi \xi \delta^{ab} f'(0) + (\rho_{1}^{ab},f_{1}),
\end{align*}  
where $\mathcal{C}^{ab}$ is an arbitrary constant and $f \in \mathcal{S}$. This solution uses the fact that any Schwartz function $f$ can be written in the form: $f(s)= f(0)f_{0}(s) + sf_{1}(s)$, where $f_{0}(0)=1$~\cite{Bogolubov_Logunov_Oksak90}. However, in order to write $\rho_{2}^{ab}$ explicitly in terms of $\rho_{1}^{ab}$ (i.e. independently of the test function $f$) the last term must be rewritable in terms of the full function $f$, and not just $f_{1}$. For the free photon case this is indeed possible because $\rho_{1}(s) = -2\pi\delta(s)$, and since $s\delta'(s) = -\delta(s)$, it follows that:
\begin{align*}
(\rho_{1},f_{1}) = -2\pi(\delta,f_{1}) = 2\pi(s\delta',f_{1})   &=2\pi(\delta',f -f(0)f_{0}) \\
&= 2\pi(\delta',f) - 2\pi f(0)(\delta',f_{0}),
\end{align*} 
which together with Eq.~(\ref{constrain}) and the constraints in Eq.~(\ref{sum_rules_rho_2}) imply that $\rho_{2}(s)= -2\pi(1- \xi)\delta'(s)$. However, for the photon or gluon propagators the form of the spectral density $\rho_{1}^{ab}$ is \textit{a priori} unknown, and so one cannot express $\rho_{2}^{ab}$, and hence the full propagator, explicitly in terms of $\rho_{1}^{ab}$. This means that a transverse-longitudinal representation as in Eq.~(\ref{prop_can}) exists for the free photon propagator [Eq.~(\ref{free_photon_prop})] but is not in general achievable for either the photon or gluon propagators. Therefore, the statement that the structure of $\widehat{D}_{\mu\nu}^{ab\, F}(p)$ has the form of Eq.~(\ref{prop_can}) due to the Slavnov-Taylor identity is evidently false. The fact that the representation of the photon and gluon propagators in Eqs.~(\ref{photon_rho2}) and~(\ref{gluon_rho2}) does not possess this form, and yet satisfies this identity, proves this point. \\

\noindent
As outlined at the end of Sec.~\ref{QCD_prop}, there are a variety of different non-perturbative techniques for analysing the structure of propagators, in particular the Schwinger-Dyson equations and lattice QFT. In light of the findings in this section, which suggest that the canonical decomposition in Eq.~(\ref{prop_can}) may no longer hold, it is important to understand if this can potentially cause inconsistencies with these techniques, and if so, whether this issue can be circumvented. In the literature it appears that in the case of both the Schwinger-Dyson~\cite{Alkofer_vonSmekal01,Alkofer_Detmold_Fischer_Maris04,Strauss_Fischer_Kellermann12,Kizilersu_Sizer_Pennington_Williams_Williams15} and lattice~\cite{Cucchieri_Mendes_Taurines05,Cucchieri_Mendes08,Dudal_Oliveira_Silva14} approaches, both the photon and gluon propagators are assumed to have the structure of Eq.~(\ref{prop_can}). In particular, in Landau gauge ($\xi=0$) it is stated that these propagators can be written: $T_{\mu\nu}D^{ab}(p^{2})$. Since the representation in Eq.~(\ref{prop_can}) is not in general achievable for either the photon or gluon propagators, extracting the structure of the propagators based on this premise could potentially lead to inconsistent results. Nevertheless, despite the failure of Eq.~(\ref{prop_can}) to hold in general, the representations in Eqs.~(\ref{photon_rho2}) and~(\ref{gluon_rho2}) \textit{are} guaranteed to hold, and this is independent of the form of the spectral densities $\rho_{1}$ and $\rho_{2}$. Moreover, if instead of calculating the propagator $\widehat{D}_{\mu\nu}^{ab\, F}(p)$ one determines the contracted quantity $g^{\mu\nu}\widehat{D}_{\mu\nu}^{ab\, F}(p)$, this representation issue no longer arises because this expression takes the form
\begin{align}
g^{\mu\nu}\widehat{D}_{\mu\nu}^{ab\, F}(p) &= 3i\int \frac{ds}{2\pi}\frac{\rho_{1}^{ab}(s)}{p^{2}-s+i\epsilon} - \frac{i\delta^{ab}\xi}{p^{2}+i\epsilon}  + \sum_{n=0}^{N+1}g_{n}^{ab}(\partial^{2})^{n}\delta(p),
\label{contract}
\end{align}
where $g_{n}^{ab}= 4c_{n}^{ab} + d_{n}^{ab}$. In the case of the photon propagator one has an analogous expression, but without the singular terms. Besides the possible singular terms, in Landau gauge one now has an expression which depends only on the spectral density $\rho_{1}$, in contrast to the non-contracted propagator. With regards to lattice calculations this means that as long as one extracts the (Euclidean) contracted propagator, one will indeed be sensitive to the behaviour of $\rho_{1}$. Similarly, by contracting the Schwinger-Dyson equation for $\widehat{D}_{\mu\nu}^{ab\, F}(p)$ with $g^{\mu\nu}$, one could in principle solve for $g^{\mu\nu}\widehat{D}_{\mu\nu}^{ab\, F}(p)$ instead of the propagator, and hence also remove the ambiguity in this case as well.

\section{Conclusions}
\label{concl}

Understanding the structure of the photon and gluon propagators is essential for probing the non-perturbative dynamics of QED and QCD. Axiomatic approaches to QFT provide a framework from which one can characterise the general properties of Lorentz covariant propagators, and the constraints imposed on them as a result of the dynamical properties of the fields in the propagators. In this paper we discuss the constraints on the photon and gluon fields, and determine the specific effect that they have on the non-perturbative structure of the photon and gluon propagators. By virtue of the abelian gauge symmetry of QED, it transpires that the photon propagator can be completely characterised by one of two different interrelated spectral densities $\rho_{1}$ and $\rho_{2}$. Moreover, in QCD the non-abelian gauge symmetry also permits additional singular terms involving derivatives of $\delta(p)$ to appear in the gluon propagator. The possibility of such terms is particularly interesting in the context of QCD, since their appearance is suggestive of confinement. Due to the distributional behaviour of the spectral densities of the photon and gluon propagators, it turns out that the lack of knowledge of these objects actually prevents one from decomposing these propagators into transverse and longitudinal components, as in the free case. Nevertheless, despite the obstruction to this decomposition both the photon and gluon propagator representations still satisfy the Slavnov-Taylor identity.

\section*{Acknowledgements}
This work was supported by the Swiss National Science Foundation under contract P2ZHP2\_168622, and by the DOE under contract DE-AC02-76SF00515.

\renewcommand*{\cite}{\vspace*{-12mm}}

\end{document}